\documentstyle[mprocl]{article}

\def\be{\begin{equation}}
\def\ee{\end{equation}}
\def\bea{\begin{eqnarray}}
\def\eea{\end{eqnarray}}

\def\line{\hbox to \hsize}    
\def\frac #1#2{{#1\over #2}}

\def\phid{\varphi^*}

\def \z{{\overline z}}

\begin{document}

\title{MAGNUS AND OTHER FORCES ON VORTICES IN SUPERFLUIDS AND SUPERCONDUCTORS}

\author{MICHAEL STONE }

\address{University of Illinois, Department of Physics\\ 1110 W. Green St.\\
Urbana, IL 61801 USA\\E-mail: m-stone5@uiuc.edu}   


\maketitle\abstracts{I discuss some of the forces acting on vortices in charged superfluids,
paying particular attention to the way that the 
Berry and Aharonov-Casher phases combine to reflect the classical magnetohydrodynamics.
}

\section{Introduction}

The motion of vortices is the only significant source of dissipation in a
superconductor, making it a topic of some technological importance.  Much of the
modern work on vortex dynamics relates to many-vortex interaction effects so it
is rather surprising to find that there is still considerable debate about the
forces acting on an individual vortex.  This debate was sparked by the claim in
papers by Ao, Thouless and Niu\cite{at,atn} that there exists a universal,
exact expression for the transverse Magnus force on a vortex.  Arising from
topological effects, the magnitude of this Magnus force should not be
affected by impurities or quasiparticles except in that they change the value of
the superfluid density.

The Ao-Thouless-Niu papers stimulated a number of authors to re-examine the
problem\cite{g,s,vano,v,kvp,mak,sonin} with conclusions
ranging between support for their viewpoint, through the suggestion 
that spectral flow, yet another topological effect, partly
cancels the Magnus force, to claims that there are no topological effects at
all.  It has also led to an ingeneous experiment aimed at a direct measurement
of the force\cite{zhu}.

Given the interests of the audience at this conference, I will focus on only one
aspect of this problem --- the effect, if any, electromagnetic interactions have
on the topological Magnus force.  In the next section I will review the
connection between the topological phase and the Magnus force for neutral
superfluids, and then show how the phase accounting is modified by
Aharonov-Casher phases due to the motion of the magnetic flux which is tied to
the vortices.  In the third section I will show how the various phase
cancellations reflect a purely classical re-routing of the momentum flux from the
charged superfluid to the background lattice.  A final discussion section will,
for completeness, briefly review how spectral flow induces relaxation effects
that need to be incuded to complete the picture.

\section {Berry and Aharonov-Casher phases}

\subsection{Neutral Fluids}

Variation of the action
\be
S=\int dtd^3x\left\{-\frac i2 (\varphi^*\partial_t\varphi -\varphi\partial_t \varphi^*)
 +\frac 1{2m}|\nabla \varphi|^2 +\frac \lambda 2 (|\varphi|^2-\rho_0)^2\right\}
\ee
gives rise to a non-linear Schrodinger equation 
\be
i\partial_t \varphi=-\frac 1{2m} \nabla^2\varphi + \lambda  (|\varphi|^2-\rho_0)\varphi
\ee
describing fluctuations of the scalar field $\varphi$ about a stationary
configuration with $|\varphi|^2=\rho_0$.  This equation is called the
Gross-Pitaevskii (GP) equation\cite{gp} and is often used as a simple model for
the motion of a Bose condensate.  Long ago Madelung\cite{madelung} showed that
the real and imaginary parts of the Schrodinger equation can be interpreted as
the equations of mass and momentum conservation for a simple fluid, so it is no
surprise that vortex solutions to the GP equation are consistent with Kelvin's
circulation theorem from elementary fluid dynamics --- i.e the vortices are
advected along with the flow.  For example, in three dimensions, there exists a
family of smoke-ring-like vortex solutions to the GP equation which move along
their symmetry axis without change of radius\cite{ich}.  Any attempt to force a
vortex to move with respect to the flow requires the application of a transverse
lift force.  In the superconductivity literature this is usually referred to as
the {\em Magnus} force\cite{nv}.

We can simultaneously obscure and illuminate the existence of this force by some
cosmetic rewriting of the action.  We begin setting $\varphi
=\sqrt{\rho}e^{i\theta}$ to get
\be
S=\int dtd^3x\left\{\rho \partial_t\theta +\frac {\rho}{2m}(\nabla\theta)^2
 +\frac{\lambda}{2}(\rho-\rho_0)^2 +\ldots\right\}
\ee
(where the dots indicate a presently  uninteresting term depending on the gradient of
$\rho$.). Next we integrate over $\rho$ to find 
(again ignoring the gradient of $\rho$ terms)
\be
S= \int dtd^3x\left\{\rho_0\partial_t\theta +\frac {\rho_0}{2m} (\nabla\theta)^2
-\frac 1{2\lambda}(\partial_t \theta +\frac 12 
mv_s^2)^2\right\},
\ee
where  $v_s=\frac 1m \nabla \theta$ is the supefluid velocity.

The first term is a total derivative and therefore does not contribute to the
classical equation of motion for $\theta$.  If we temporarily suppress the
$v_s^2$ part of the third term, variation of the action gives rise to a wave
equation which  describes sound waves propagating in the fluid.  We
should not, however, use this truncated equation for the vortex dynamics.
Wave-equation dynamics for an angular variable like $\theta$ appears in the
relativistic form of the $X-Y$ model and it is well known that in such a model a
vortex loop will rapidly shrink and disappear\cite{neu}.  Throwing away the higher
order terms in $v_s$, although common in the superconductivity
literature\cite{suhl}, is therefore not safe when one is interested in vortex
motion.  This because omitting these terms destroys galilean
invariance\cite{witten}.

It is not safe to discard the first term $\rho_0\partial_t\theta$
either\footnote{This is done, for example, in the otherwise excellent texbook by
Popov \cite{popov}.}  If one wishes to use $S$ to compute the partition function
via an imaginary time path integral over configurations with periodicity
$\beta=1/kT$, the first term will give rise to a topological
phase.  Although the field $\varphi$ must return to its original value at $\tau
=\beta$, the angle $\theta$ will in general have wound through $2\pi N$ where
$N$ is an integer.  In two dimensions the order parameter phase associated with
a vortex contains a part $\theta({\bf r},t) = {\rm Arg}({\bf r}-{\bf r}_L(t))$
therefore moving the vortex in a closed path will contribute a phase factor
equal to
\be
\exp \{ 2\pi i\rho_0(Area)\} 
 \ee
where $(Area)$ is the area enclosed by the vortex trajectory.  This is exactly the same
phase factor that would be generated by a charged particle with interaction
 action $\oint  e\dot x^\mu A_\mu\, dt$ moving in a uniform magnetic field.  It
 must therefore signal the existence a transverse ``Lorentz'' force of magnitude
 $2\pi \rho_0 V_{vortex}$.  Taking into account that the circulation around the
 vortex is $\kappa= 2\pi/m$ this is exactly the expected magnitude of the
 classical Magnus force $\rho_0 m \kappa V_{vortex}$.
 
This connection between this topological ``Berry'' phase and the Magnus force in
a superfluid was pointed out in Ref.1. It became the motivation for a more
general theorem on the existence of transverse forces on vortices\cite{atn}.

\subsection{Charged Fluids}

What happens to the vortex-induced Berry phase when the fluid is composed of
charged particles?  To find out it is simplest to perform a duality map which
raises the status of the vortices from mere defects in the order parameter to
the central objects of discussion.  There are many references for such maps, but
since the notation and inspiration for this section comes directly from a
paper\cite{lk} written by a member of the audience I will recommend that for an
introduction.

For geometric simplicity I will restrict myself to two dimensions.  In the
following roman subscripts such as $a$ will run over the two space dimensions
while greek subscripts such as $\alpha$ will run over both euclidean time
($\alpha=0$) and space dimensions.

When we couple our scalar field to a gauge field the Lagrange density becomes
\be
L= \frac 12 (\phid(\partial_0-ieA_0)\phi- \phi(\partial_0 +ieA_0)\phid)
 +\frac 1{2m}|(\partial_a-ieA_a)\phi|^2+\frac \lambda 2(|\varphi|^2-\rho_0)^2 +\frac 14 F^2.
\label{eq:Lcharge}
\ee
Set $\varphi=fe^{i\theta}$
to write this as 
\be
L=if^2(\partial_0\theta-eA_0)+ \frac 1{2m}(\partial_a f^2)^2 +
\frac {f^2}{2m}(\partial_a\theta-eA_a)^2
+\frac \lambda 2(f^2-\rho_0)^2 +\frac 14 F^2.
\ee
Now introduce  a pair of Hubbard-Stratovich fields $C_a$ and also set $C_0=f^2$ to get
\be
L\to L'= iC_\mu(\partial_\mu\theta
-eA_\mu) +\frac 1{8m} \frac {(\partial_a C_0)^2}{C_0} +
\frac {m}{2C_0} C_a^2 +\frac \lambda 2(C_0-\rho)^2 +\frac 14 F^2.
\ee

Next we isolate the vortex part of the phase $\theta$ by writing
$\theta=\bar\theta+\eta$ where $\bar \theta=\sum_i {\rm Arg}({\bf r}-{\bf
r}_i(t))$ is the singular part of the phase due to vortices at ${\bf r}_i$, while
$\eta$ is the remaining non-singular part.  Integration over $\eta$ enforces the
mass conservation equation $\partial_\tau C_0+\partial_a C_a=0$. This is
automatically satisfied if we write
\be
C_\mu=\epsilon_{\mu\nu\sigma}\partial_\nu b_\sigma.
\ee
Regarding $b_\sigma$ as a gauge potential then suggests renaming 
\be
C_a\to \tilde E_a \qquad C_0=\tilde H,
\ee
a set of pseudo electric and magnetic fields. After defining 
\be
K_\mu=\epsilon_{\mu\nu\sigma}\partial_\nu\partial_\sigma \bar \theta
\ee
to be the vortex 3-current density we find that the Lagrange density takes the form
\be
L'= ib_\mu(K_\mu-e\epsilon_{\mu\nu\sigma}\partial_\nu A_\sigma) +
\frac 1{2\rho} |\tilde E|^2 +\frac 12 (\tilde H-\rho_0)^2 + \frac 14 F^2 +\ldots
\ee
where the dots represent the same higher order gradients of the density
that were omitted in the earlier section.
Assuming $\rho\approx \rho_0$, we can  write the Lagrange density as
\be
L'= ib_\mu(K_\mu-eF^*_\mu) +\frac 1{2\rho_0} |\tilde E|^2 +\frac 12 (\tilde H-\rho_0)^2
 +\frac 14 F^2
\ee
where $F^*_\mu= \frac 12 \epsilon_{\mu\nu\sigma} F^{\nu\sigma}$ is the dual of
the real electromagnetic flux.

We see that the vortex current is coupled to a gauge field $\tilde E$ that
apparently mediates coulomb interactions between vortices but actually accounts
for the attractive and repulsive forces due to the kinetic energy of the fluid,
and to a magnetic field $\tilde H$ which has an equilibrium value $\rho_0$ even
when the vortex is at rest.  Motion with respect to this background field gives
rise to a Lorentz force which is just the Magnus force in this pseudo-field
language.  We also see that the source for the fields is not the bare vortex
current but the vortex current less the flux of the genuine magnetic field that
acompanies the vortex core.  As in any Abrikosov flux tube solution, this
magnetic flux completely screens the vortex current and appears to eliminate
both the ``coulomb'' interaction between the vortcies {\em and} the Magnus
force.  The coulomb screening is a geniune effect -- the dual of the
Anderson-Higgs mechanism.  Is the Magnus force banished as well?  The answer is
no!  We have omitted an important term in Eq.\ref{eq:Lcharge}.  This is the
interaction of the electromagetic field with the background ions.
Consideration of this is necessary if we are not to have infinite coulomb energy.
Including an $L_{ions}=ie\rho_{ion} A_0$ of the standard form and writing the
ionic 3-current (a static charge of course) as the curl of a field $b_\mu^{ion}$
the final form of the vortex field interaction becomes
\be
ib_\mu(K_\mu-eF^*_\mu) + ieb_\mu^{ion}F^*_\mu +\frac 1{2\rho_0} |\tilde E|^2
 +\frac 12 (\tilde H-\rho_0)^2+\frac 14 F^2.
\ee
When the fluid is at rest with respect to the background charges away from the
vortices (which it has to be unless one is within a penetration depth of the
boundary of the system) the two terms prortional to $F^*_\mu $ cancel and the
Magnus force is restored.

In terms of topological phases, what is happening  is that the Berry phase
from the superfluid is being supplemented by two Aharonov-Casher phases\cite{ac}
arising from moving the magnetic flux line round the charges in the system.  One
Aharonov-Casher phase comes from moving the flux through the mobile fluid
charges and one from moving it through the static background ion charges.
Coulomb interactions force the net charge inside any macroscopic region to be
zero so the two Aharonov-Casher phases must cancel.

The results of this section differ from those in  some recent
papers\cite{vano}. Although in the end they obtain the same total
force on the vortex, the authors of these papers claim that inclusion
of electromagnetic interactions causes the topological phase terms in
the effective action to cancel.  This cancellation originates in  the
form they chose for  the interaction with the background charge.  I
have chosen to represent it as $ie\rho_{ion} A_0$ whereas these
papers argue that  gauge invariance requires us to write this
interaction term as a gauge covariant derivative $ie\rho_{ion}
(A_0-\partial_0\theta)$, where $\theta$ is the phase of the {\em
electron} order parameter.  Then, because $\rho_{ion}=\rho_0$ the
terms in $\partial_0\theta$ cancel.  I would argue that since the
background ion current is conserved separately from the electron
current, the conventional form is gauge invariant without any
additional term.  Similar opinions have been expressed  by   Zhu,
Gaitan and Volovik\cite{zhu2}.  In the next section I will argue that
my choice of interaction is the correct one by relating the phase
cancellations to the actual mechanical forces acting on the system.

\section{ Kutta-Joukowski Theorem for Charged Fluids}

The original example\cite{ac} exhibiting the Aharonov-Chasher phase was
constructed to be similar to the Bohm-Aharonov effect in that there was a
quantum effect even though the classical force acting on the moving object vanished.
However, just as the Bohm-Aharonov phase in  a region with non vanishing
magnetic field becomes path dependent and signals the existence of the Lorentz
force, so the Aharonov-Casher phase in a region of non-vanishing charge density
becomes path dependent and signals that moving flux through a charge
distribution requires application of transverse force --- the reaction force to
the Lorentz force the charges feel in the moving magnetic field.  (More
accurately the charges feel the electric field arising from the
Lorentz-transformed magnetic field).  We can therefore avoid all controversy
over cancelling phases and gauge invariance by focussing on the equivalent
classical forces experienced by the fluid and the background ions.

\subsection{Neutral Fluids}

To understand the forces on the vortex, or equivalently the momentum flux into
and out of the vortex core, we may as well look at the forces on a hypothetical
solid body occupying the same region as the core.  We will begin by reviewing
the traditional derivation\cite{lamb} of the lift force on an aerofoil with
circulation $\kappa$ which is being held stationary in electrically neutral,
two-dimensional, incompressible potential flow.

This force is most easily obtained from the momentum flux tensor. 
\be
T_{ij}=\rho m v_i v_j + g_{ij} P.
\ee

Since we are interested in steady, irrotational motion with constant density we
may use Bernoulli's theorem, $P+\frac 12 \rho m |v|^2= const.$, to substitute
$-\frac 12 \rho m|v|^2$ in place of $P$.  (the constant will not affect the
momentum flow).  With this substitution $T_{ij}$ becomes a traceless symmetric
tensor
\be 
T_{ij}=\rho m(v_i v_j -\frac 12 g_{ij}|v|^2).
\ee

In two dimensions it is convenient to use complex coordinates $z=x+iy$ where the
metric tensor becomes $g_{zz}= g_{\z \z}=0$, $g_{z \z}= g_{\z z}=\frac 12$ and
$g^{z \z}=g^{\z z}=2$ etc.  In these coordinates $v_z=\frac 12 (v_x-iv_y)$ and
\be
T\equiv T_{zz}= \frac 14(T_{xx}-T_{yy} -2i T_{xy})= \rho m (v_z)^2.
\ee
$T$ is the only component of $T_{ij}$ we will need to consider.  $T_{\z \z}$ is simply 
$\overline T $ while $T_{z \z}=0=T_{\z z}$ because $T_{ij}$ is traceless.

Any body-force $f_i$ acting on  the fluid is given by the divergence of the
momentum flux tensor
\be
f_i = g^{kl}\partial_k T_{l i}
\ee
or, in our coordinates,
\be
f_z= g^{\z z}\partial_\z T_{zz}+ g^{z \z}\partial_z T_{\z z} = 2 \partial_\z T.
\ee
From this we see that, in a steady flow, 
the momentum flux $\dot P$ out of a region $\Omega$ must be given by 
\be
\dot P_z= \frac 1 {2i} \int f_z\, {d\z dz}  = \frac 1i \int \partial_\z T\,{d\z dz}=
\frac 1i \oint_{\partial \Omega} T \,dz.
\label{eq:force}
\ee
When there is no body-force, $T$ is analytic and the integral will be
independent of the choice of contour $\partial \Omega$.

To apply this result to our aerofoil we should take $\partial \Omega$ to be its
boundary.  Then $\dot P_z$ becomes minus the integral of the pressure
over the boundary of the body,   i.e  the force exerted on the fluid by the
aerofoil.  Evaluating the integral is not immediately possible because the
velocity $V_z$ will be a complicated function of the shape of the body.  We can,
however, exploit the contour independence of the integral and evaluate the
integral over a contour encircling the aerofoil at large distance where the
flow field takes the form
\be
v_z=U_z + \frac {\kappa}{4 \pi i}\frac 1 z +O(\frac 1{z^2}).
\ee
To confirm that  this flow has the correct circulation  we  compute  
\be
\oint {\bf v}\cdot d{\bf r} =\oint v_z dz+ \oint v_\z \,d\z =\kappa.
\ee

Subsituting in Eq.\ref{eq:force} we find that
\be
\dot P_z= -i \rho m \kappa U_z.
\ee
In conventional coordinates the reaction force  on the body ${\bf F}=-\dot{\bf  P}$ is
\bea
F_x&= &\rho m \kappa U_y\nonumber\\
F_y& = - &\rho m \kappa U_x.
\eea
The body therefore exerts a transverse lift force on the fluid proportional to the
product of the circulation with the asymptotic velocity.  This is the
Kutta-Joukowski theorem.

\subsection{Charged Fluids}

How is this result modified when the neutral fluid is replaced by one composed
of charged particles?  We wish to model a charged superfluid where the
superfluid velocity is obtained from the order parameter phase, $\theta$, via
the equation
\be
{\bf v}=\frac 1m (\nabla\theta -e {\bf A}).
\ee
The flow is therefore  no longer irrotational but instead has vorticity
\be
\omega= \nabla \wedge {\bf v} =-\frac em \nabla\wedge {\bf A} = -\frac em {\bf B}.
\ee

The relation $e{\bf B} + m \omega =0$ gives rise to the Meissner effect:
using the Maxwell equation $\nabla\wedge {\bf H}= {\bf J}$ with ${\bf J}=\rho e
{\bf v}$ we find
\be
\nabla^2 {\bf B} + \mu_0 \frac {\rho e^2}{m} {\bf B}=0.
\label{eq:meissner}
\ee

Eq.\ref{eq:meissner} implies that both the magnetic flux and the vorticity decay
 exponentially away from a flux-creating disturbance.  A vortex defect is such a
 disturbance.  The $2\pi$ phase winding produces a solenoid-like current near
 the vortex core.  This current creates a magnetic field which in turn induces a
 local vorticity that tends to screen that of the defect over the penetration
 length $\lambda_0 = \sqrt{ m/ (\rho e^2 \mu_0)}$.  The circulation about the
 vortex is therefore path-dependent and tends to zero as the integration path
 becomes further from the vortex core.

From the fact that the induced vorticity exactly screens the circulation round the
 bare vortex we can find the  net flux 
\be
\Phi= \int B_3 \,dx dy = - \frac m e \int \omega \,dxdy = \frac m e \frac {2\pi}{m}
 = \frac {2\pi}{e}.
\ee
For the Cooper pairs of a BCS superfluid $m= 2\times$ (mass of electron) and $e
= 2\times$ (charge of electron), so this flux quantum is half that appearing
in the quantum Hall effect.

As before, we must not forget the charge of the background ions.  Without a
background charge $\rho_{ion}\approx \rho$ there would be a huge Coulomb field.
Indeed any deviation from equality between the two charges will tend to zero
over a Debye screening length so we will assume that $\rho_{ion}= \rho$
everywhere.  Further any tendency of the superfluid to flow relative to this
background charge will create a magnetic field and the resulting Meissner effect
will restrict  the flow to within a penetration length $\lambda_0$ of the
boundary of the system.  To avoid this current inhomogeneity and to obtain a
theorem analagous to that of Kutta-Joukowski we will therefore assume that the
asymptotic fluid flow $U$ is equal to that of the rigid motion background ions.
This means that we are working in the rest frame of the vortex which is itself
moving at a steady velocity $-U$ with respect to both the asymptotic fluid and
the background ion crystal.  We also assume that our body $\Omega$ contains
static charge with density equal to that of the surrounding fluid.

Despite the non-zero vorticity it is easy to see that the momentum flux tensor
remains $T_{ij}=\rho m(v_i v_j -\frac 12 g_{ij}|v|^2)$.  The momentum supplied by the 
aerofoil is therefore still given by
\be
\dot P_z=\frac 1i \oint_{\partial \Omega} T \,dz,
\ee
but now $T$ is no longer analytic.  Indeed we have that
\be
\partial_\z T = \frac i 2 e \rho B v_z.
\ee
We {\it must} therefore take $\partial \Omega$
as the boundary of the aerofoil. 
(From now on we write $B$ for $B_3$ since this is the only non-zero
component of the field).  Although we cannot use analyticity to send the
integration contour off to infinity, we can still integrate by parts and find
\be
\frac 1i \oint_{\partial \Omega} T \,dz=
 - \frac 1i \int_{{\bf R}^2 \backslash\Omega} \partial_\z T \,d\z dz
\ee 
\be
= -\frac {e\rho}{2}\int_{{\bf R}^2 \backslash\Omega} B v_z \,d\z dz.
\ee
This last expression is simply the Lorentz force on the fluid outside the aerofoil.

We now note that the Maxwell equation $\nabla \wedge H= J$  is in our complex
coordinates
\bea
i\partial_z B &= e\mu_0\rho (v_z -U_z)  &\quad {\rm in}  
\quad {\bf R}^2 \backslash\Omega\nonumber\\
&= -e\mu_0\rho U_z &\quad {\rm in} \quad \Omega.
\eea
So 
\be
\dot P_z = -\frac i{2\mu_0} \int_{{\bf R}^2 \backslash\Omega} (B\partial_z B)\,d\z dz
-\frac {e\rho}{2} \int_{{\bf R}^2
 \backslash\Omega} B U_z \,d\z dz.
\label{eq:Uint}
\ee

The last term in Eq.\ref{eq:Uint} is independent of the flow field outside the
body but still depends on details of the system through the domain of
integration.  We will see that we can combine both terms to get an expression
for the lift that is completely independent of any such details.

We observe that 
\bea
  -\frac i{2\mu_0} \int_{{\bf R}^2 \backslash\Omega} (B\partial_z B)\,d\z dz& =
 & -\frac i{2\mu_0} \int_{{\bf R}^2 \backslash\Omega}
  \partial_z \frac 12 B^2\,d\z dz\nonumber\\
& = &- \frac i{2\mu_0} \oint_{\partial \Omega} \frac 12 B^2 \,d\z\nonumber\\
& = & -\frac i{2\mu_0} \int_{\Omega} \partial_z  \frac 12 B^2 \,dz d\z \nonumber\\ 
& = & \frac 12 \int_{\Omega} e\rho B U_z\, dz d\z.
\eea

Putting the two parts together then gives
\be
\dot P_z=\frac 12 e \rho \int_{{\bf R}^2} B U_z \,dz d\z = -i e\rho \Phi U_z
\ee
This expression for the lift force on the aerofoil depends only on the total
flux and the asymptotic current.  Taking into account the relationship between
the neutral-fluid quantum of circulation and the flux in the Abrikosov vortex we
see that the force on the aerofoil, or equivalently the momentum flux into or
out of a vortex, is unchanged by the introduction of electromagnetic
interactions.

What is happening  is that in the neutral case momentum is flowing out of
the vortex and increasing the net momentum of the fluid.  In the charged case
the moving vortex supplies exactly the same amount of momentum to the fluid, but
the magnetic field acts as a clutch coupling the fluid to the background
ions.  Any relative motion  creates a magnetic field that tends to inhibit
it.  The momentum is therefore transfered to the background ions.  That is why
the force the aerofoil exerts on the fluid is exactly equal to the Lorentz force
the magnetic field exerts on the background ions as a result of their being
dragged through the vortex's flux tube.  A similar mechanism occurs in the
quantum Hall effect where motion of either a Laughlin quasiparticle or a
skyrmion transfers momentum to the electrons, and they in turn hand it on to the
magnet\cite{stone1}.  An alternative way of thinking of the same physics is to
follow Nozieres and Vinen\cite{nv} and realise that in the charged fluid the
circulation, and hence the Magnus force, is reduced as we look at contours at
larger and larger distance from the vortex.  The lost Magnus force is however
exactly compensated by the Lorentz force on the fluid within the contours.
This re-routing of the momentum exactly mirrors the phase cancellations described in
 section 2.

\section{Discussion}

We have treated the superfluid as if it were a Bose condensate of Cooper pairs
and therefore modeled its motion by the GP equation.  This model differs in
several ways from BCS electrons in a real metal.  Firstly the GP equation is
galilean invariant.  This, in my opinion, is not a problem.  One often models
electrons in a metal by taking a parabolic band dispersion curve $\epsilon
=k^2/2m^*$, and the resulting pseudo-galilean invariance provides a useful
check on one's computation. 

Secondly some people object to modeling the dynamics by a first-order equation
at all.  For the neutral case this should not be an issue.  It has been shown by
various authors\cite{shakel,stone2,thouless} that this is the correct form of
the low energy dynamics for a galilean invariant BCS system at $T=0$.  The
quantity $\varphi$ appearing in the GP equation is not, however, the BCS order
parameter.  It is instead the combination $\varphi=\sqrt{\rho}e^{i\theta}$ where
$\theta$ is the phase of the order parameter and $\rho$ is the fluid density.
(You may have noticed that throughout this talk I have never refered to
the scalar field $\varphi$ as the ``order parameter''.  I have only refered its
phase as the ``phase of the order parameter''!)  These variables are those
associated with the low-energy Goldstone mode.  The magnitude of the BCS order
parameter $|\Delta|$ is not associated with  Goldstone-mode dynamics and so
does not appear in the low-energy effective action.

Finally, and most importantly, BCS vortices have internal dynamics associated
with quasiparticle core states.  Volovik suggested that there should be a novel
axial-anomaly driven spectral flow among these states in a moving vortex, and
that this spectral flow would cancel most of the Magnus force\cite{v}.  It turns
out that once one takes the discreteness of the core states into account then
one must include relaxation processes in order to see their
effect\cite{stone3,kvp}.  With relaxation  included  this turns out to be
a previously known effect known as the Kopnin-Kravtsov (KK) force\cite{kk}, a
sort of transverse friction which exhanges momentum with the crystal lattice by
scattering of core quasiparticles off lattice defects.  How does this extra
transverse force fit in with the claims of Ao et al.~that the Magnus force is
all there is?  The theorem of Ao-Thouless and Niu\cite{atn} states that if a
vortex moves against the flow, then momentum enters the superfluid at exactly
the rate given by the Magnus force.  When we include the KK friction a vortex
placed in a flow field {\em will} find itself moving at a velocity different
from the asymptotic flow.  Momentum must therefore enter the fluid at exactly
the rate given by the theorem\cite{atn}.  In order to solve for the relative
speed of the vortex and the fluid, though, we need {\em another} expression for
the rate at which momentum is entering the system.  The actual mechanism by
which the momentum is being gained is that momemtum is being transfered to the
core states from the lattice by KK scattering, and from there flows out into the
fluid in accord with Eq.~\ref{eq:force}.  The rate of KK scattering depends on
the relative velocity of the vortex and the crystal lattice, so equating these
two expressions for the rate of momentum transfer  determines
  the steady state vortex
velocity in terms of the relative velocity of the lattice and the flow field.

\vspace*{-2pt}
\section*{Acknowledgments}

This talk describes work partly carried out at the Institute for Theoretical Physics at
UC Santa Barbara.  Work at the ITP was supported by the National Science
Foundation under Grant PHY94-07194.  Work at Illinois was supported under
Grant number DMR94-24511.  I would like to thank Ping Ao, David Thouless, Carlos
Wexler, Frank Gaitan, Anne van Otterlo and Xiaomei Zhu for valuable
conversations.

\eject

\end{document}